\documentclass{article}
\topmargin = - 0.5 cm
\newcommand{\Gindex}{\mbox{$G$-index}}
\def\Dslash{\setbox0=\hbox{$D$}D\hskip-\wd0\hbox to\wd0{\hss\sl/\/\hss}}
\newcommand{\DS}{\Dslash}
\usepackage{amsmath}
\usepackage{amssymb}

\newcommand{\GSC}{Guillemin--Sternberg conjecture}

 \newcommand{\begeq}{\begin{equation}}
\newcommand{\bea}{\begin{eqnarray}}
\newcommand{\eea}{\end{eqnarray}} \newcommand{\nn}{\nonumber}

\newcommand{\ca}{$C^*$-algebra} 
 \newcommand{\rep}{representation}

\newcommand{\sthom}{$\mbox{}^*$-homomorphism}

 \newcommand{\til}{\tilde}
\newcommand{\raw}{\rightarrow}

\newcommand{\law}{\leftarrow}

\newcommand{\ot}{\otimes} 
\newcommand{\la}{\langle} \newcommand{\ra}{\rangle}

\newcommand{\x}{\times} 
 
\newcommand{\cin}{C^{\infty}}

\newcommand{\inv}{^{-1}}

\newcommand{\spinc}{\mathrm{Spin}^c}
 \newcommand{\Gm}{\Gamma}
 \newcommand{\Dl}{\Delta}

 \newcommand{\sg}{\sigma}
  
 \newcommand{\phv}{\varphi}
 \newcommand{\ps}{\psi} 
\newcommand{\om}{\omega} 

 \newcommand{\g}{\mathfrak{g}}

\newcommand{\CJ}{{\mathcal J}}

 \newcommand{\CS}{{\mathcal S}}

\newcommand{\C}{{\mathbb C}} 
 
 \newcommand{\R}{{\mathbb R}}
 \newcommand{\Z}{{\mathbb Z}}
  \makeatletter
\newskip\tempskip \def\endproof{{\parfillskip24\p@ plus\@ne
fil\@@par}\tempskip\prevdepth
\ifdim\lastskip=\z@\tempskip\z@\else\vskip-\lastskip
\ifdim\tempskip>4\p@ \tempskip.5\tempskip \else \tempskip\z@\fi\fi
\nobreak\vskip-\baselineskip\vskip-\tempskip\noindent\hbox
to\hsize{\hfill
$\blacksquare$}\par\vskip\tempskip\vskip\abovedisplayskip\@doendpe}
\makeatother \makeatletter
\newskip\tempskip \def\endiproof{{\parfillskip24\p@ plus\@ne
fil\@@par}\tempskip\prevdepth
\ifdim\lastskip=\z@\tempskip\z@\else\vskip-\lastskip
\ifdim\tempskip>4\p@ \tempskip.5\tempskip \else \tempskip\z@\fi\fi
\nobreak\vskip-\baselineskip\vskip-\tempskip\noindent\hbox
to\hsize{\hfill
$\Box$}\par\vskip\tempskip\vskip\abovedisplayskip\@doendpe}
\makeatother 

\newcommand{\otc}{\circledcirc}

\newcommand{\car}{\circlearrowright}
\begin{document}
\thispagestyle{empty}
\title{Functorial quantization and \\ the Guillemin--Sternberg conjecture\thanks{Dedicated to Alan Weinstein, at his 60th birthday. To appear in S.T. Ali et al (eds.), Proc.\ XXth Workshop on Geometric Methods in Physics, Bialowieza, 2002. }}
\author{\textbf{N.P. Landsman} \\ \mbox{} \hfill \\
Korteweg--de Vries Institute for Mathematics\\
University of Amsterdam\\
Plantage Muidergracht 24\
NL-1018 TV AMSTERDAM\\ THE NETHERLANDS\\
\mbox{} \hfill \\
email \texttt{npl@science.uva.nl}}
\date{\today}
\maketitle
\begin{abstract}
 We propose that geometric quantization of symplectic manifolds is the arrow part of a functor, whose object part is deformation quantization of Poisson manifolds.
The `quantization commutes with reduction' conjecture of Guillemin and Sternberg 
then becomes a special case of the functoriality of quantization. In fact, our formulation yields almost unlimited generalizations of the Guillemin--Sternberg conjecture, extending it, for example, to arbitrary Lie groups or even Lie groupoids.
Technically, this involves symplectic reduction and Weinstein's dual pairs on the classical side, and Kasparov's bivariant K-theory for $C^*$-algebras (KK-theory) on the quantum side.\end{abstract}
\newpage
 \section{Introduction}
 The theory of constraints and reduction in mechanics and field theory is 
important for physics, because the fundamental theories describing Nature
(viz.\ electrodynamics,  Yang--Mills theory, general relativity, and possibly also string theory)  are a priori formulated as constrained systems (cf.\ \cite{Sun}).
 The systematic investigation of classical constrained  systems was initiated by Dirac, whose  ideas were reformulated  mathematically as the theory of symplectic reduction
(see, e.g., \cite{BSF,NPL}). The procedure known as 
Marsden--Weinstein reduction \cite{AM,MW,Mey} is a special case of this theory, which is  easy to formulate, yet very rich in mathematical and physical applications.
According to this procedure, a suitable action $G\car M$ of a Lie group $G$ on a symplectic manifold $M$ produces another symplectic manifold, the reduced space $M^0$, which is a certain subspace of $M/G$. See \cite{MW2} for a recent overview.

In general, the traditional idea of quantization has always been that a 
phase space, i.e., a symplectic space
$M$,  should be quantized by a Hilbert space $H(M)$, and that the classical observables, viz.\ the (real-valued) smooth functions on $M$ should be quantized by (self-adjoint) operators on $H$, which after all play the role of observables in quantum theory.  What is the relationship between the quantization of $M$
and the quantization of the reduced space $M^0$?

The quantization of constrained systems was first analyzed in a general setting 
in  \cite{Dir64}, but there still exists no complete and satisfactory mathematical theory. Given some notion of quantization $Q$, the basic problem in such a theory would be to formulate a possible quantum analogue of the classical reduction procedure $R$, and compare the result of applying this procedure to the quantization of the unconstrained classical systems with  the quantization of the classically reduced system. 
One would then hope that the order of quantization and reduction does not matter; this hope is  symbolically expressed by `$[Q,R]=0$.'  This `quantization commutes with reduction' principle can be turned into a mathematical conjecture once a precise meaning has been assigned to the operations $Q$ and $R$. See \cite{GGK}  for a survey of the literature on this problem in the context of geometric quantization, and cf.\  \cite{NPL} for references (pre 1998) on other approaches.

 In the context of (what we now call) Marsden--Weinstein reduction, Dirac proposed that the  $G$ action on $M$ should be quantized by  a unitary \rep\ $U$ of $G$ on $H(M)$, while  the so-called weak observables act on $H(M)$  by operators commuting with $U(G)$.
The quantized reduction operation $RQ$ then consists in taking the $G$ invariant part $H(M)^G$ of $H(M)$, on which the weak observables then act by restriction. This idea makes rigorous mathematical sense in general only when $G$ is compact. If, in addition,  $M$ is compact, one expects $H(M)$ to be finite-dimensional, and similarly for $H(M^0)$, so that the weakest possible form of the `$[Q,R]=0$' conjecture, in which the action of observables is ignored, would be 
 \begin{equation}
 H(M)^G\cong H(M^0). \label{QR0}
\end{equation}
Here the $\cong$ sign stands for unitary isomorphism, and since the dimension is the only such invariant of a Hilbert space,  one really is talking about a simple equality between numbers, i.e.,  $\dim(H(M)^G)=\dim(H(M^0))$.

Despite various refinements \cite{GGK}, some of which will be discussed below, (\ref{QR0})  is basically the form in which the conjecture has been studied in the mathematical literature. This literature started with the seminal paper \cite{GS82}, after which the conjecture in any form
resembling (\ref{QR0}) is usually named. 
Using geometric quantization, they proved (\ref{QR0}) under certain assumptions,
among which the compactness of $M$ and $G$ are crucial. 
It is hard to think of a more favourable situation for quantization theory then the one assumed in \cite{GS82}. Partly in order to generalize the Guillemin--Sternberg conjecture, in the mid-1990s a novel notion of quantization came up, which seems to incorporate  all good features of geometric quantization whilst circumventing a number  of its pitfalls; see \cite{GGK,Sja}, and references therein.  
This definition of quantization is sometimes attributed  to Raoul Bott.

 In this approach, quantization is  simply defined as the index of a suitable 
Dirac operator $\DS$  naturally associated to $M$; when $M$ carries a $G$ action $G\car M$, this index is understood in the equivariant sense, so that the quantization of $M$, or rather of $G\car M$, is an element of the \rep\ ring $R(G)$ of $G$. The quantization of the reduced space $M^0$ remains an integer. Taking the $G$ invariant part of a \rep\ induces a map $R(G)\raw\Z$, in terms of which the Guillemin--Sternberg conjecture can then be stated in a very elegant form. In that  form, it was proved  in \cite{Mei,MS}; also see \cite{GGK,Par1} for other proofs and  further references.

These ideas still  only apply  to the situation where $M$ and $G$ are compact
(though cf.\ \cite{Par2} for a special case where at least $M$ is noncompact), which is
 highly undesirable for applications to both physics and mathematics.
Furthermore, it would be welcome to have some direct motivation for 
the  notion of quantization as an (equivariant) index, and if possible also 
to incorporate some extra structure. For example, when no $G$ action is around, 
Bott's definition of quantization  merely produces a number, and the entire idea of quantizing functions on $M$ by operators is lost. 

These problems can be addressed by combining geometric quantization with
 deformation quantization. In the latter, a Poisson manifold is quantized
by an associative algebra, subject to a number of conditions.
 In the `formal' setting, this should be an algebra over the commutative ring
$\mathbb{C}[[\hbar]]$ of formal power series in one real variable \cite{BFFLS}, 
whereas in the `strict' setting this should be a \ca\ over the commutative \ca\ $C(I)$ of continuous functions on the interval $I=[0,1]$ \cite{NPLQF}. As in the entire context of relating classical to quantum mechanics \cite{NPL}, the language of \ca s is particularly attractive here. For our present purposes, it is sufficient to work with ordinary \ca s (instead of \ca s  over $C(I)$); this amounts to quantizing at a fixed value of $\hbar$, as is usual also in geometric quantization. 
This simplification entails the need to impose prequantizability conditions on the symplectic manifolds in question. 
 
In \cite{NPLQF} we proposed that quantization should be seen as a functor between categories whose arrows are equivalence classes of bimodules. What this means is rather different in the classical and in the quantum case \cite{NPLQF,OBWF}.
In the former, the arrows
between Poisson manifolds are isomorphism  classes of  
 symplectic  dual pairs \cite{K1,W1}. In the latter, the arrows between (separable) \ca s are homotopy  classes of  Kasparov  bimodules  \cite{Kas1}. Such bimodules are generalized Hilbert spaces equipped with a generalized Fredholm operator, such as (a bounded version of) a Dirac operator $\DS$. 

In other words, quantization should map (isomorphism  classes of)
symplectic dual pairs into (homotopy  classes of)  Kasparov  bimodules. More precisely, 
 if Poisson manifolds $P_1$ and $P_2$ are quantized by
(separable) \ca s $Q(P_1)$ and $Q(P_2)$, respectively, then a symplectic dual pair
$P_1\law M\raw P_2$ should be quantized by an element of 
the Kasparov group $KK(Q(P_1),Q(P_2))$. 
In the special case of a symplectic  dual pair $pt\law M\raw pt$,  quantization should therefore
produce an element of $KK\C,\C)\cong\Z$, i.e., an integer.  This is precisely what
 Bott's index-theoretic definiton of quantization does. 

In \cite{NPLQF} we had no idea what the quantization functor should look like, and therefore missed the connection between the envisaged functoriality of quantization and the Guillemin--Sternberg conjecture.   We now propose that a suitable  generalization of Bott's definiton of quantization will do the job, and will check that
 the Guillemin--Sternberg conjecture is actually a special case  of functoriality. 
Conversely, requiring the functoriality of quantization on suitable symplectic  
dual pairs leads  to almost unlimited generalizations of the Guillemin--Sternberg conjecture. 
For example, one can now  remove the restriction that   $M$ and $G$ 
have to be compact, in which case the  quantization functor  constructs  the quantization of a canonical $G$ action $G\car M$ as a generalized equivariant index as defined in the K-theory of group \ca s \cite{Con}. This relates the Guillemin--Sternberg conjecture to the Baum--Connes conjecture in noncommutative geometry \cite{BCH,Con}, in which it is postulated  that
the K-theory of a group \ca\ is exhausted by such indices. Moreover, 
techniques that have been developed in the context of the Baum--Connes conjecture 
\cite{Con,LeGall,Pat} enable one to state a generalized Guillemin--Sternberg conjecture even for Lie groupoid actions. Finally, our approach incorporates and illuminates the use of shriek maps in K-theory \cite{AS1,C82,Con,CS,HS}, whose functoriality turns out to be  a special case of the
functoriality of quantization. 

Since this paper relates two different areas of mathematics, we have tried to make
it largely self-contained. Following a brief review of 
 classical reduction, we  recall the  idea of looking  at
symplectic  dual pairs as arrows between Poisson manifolds. We then
review the Guillemin--Sternberg conjecture in its original form, and subsequently,
following a recapitulation of 
$\spinc$ structures and Dirac operators, in its  modern form based on 
Bott's definition of quantization. We then explain how the quantization context
naturally leads to KK-theory, including  the idea of interpreting homotopy classes
of  Kasparov bimodules as arrows between \ca s.  
We then show that the  Guillemin--Sternberg conjecture is a special case of the
functoriality of quantization. 
In the final two sections  we consider generalizations of the  Guillemin--Sternberg conjecture 
by relating  quantization to  K-homology and to foliation theory, respectively. 
 \section{Classical reduction}
A Poisson manifold $M$ is a manifold equipped with a Lie bracket $\{\, ,\,\}$ on $\cin(M)$
with the property that for each $f\in\cin(M)$ the map $g\mapsto \{f,g\}$ defines 
a derivation of  the commutative algebra structure of $\cin(M)$ given by
pointwise multiplication.  Hence this map is given by a vector field $\xi_f$, called
the Hamiltonian vector field of $f$.  Symplectic manifolds are special instances of Poisson manifolds, characterized by the property that the  Hamiltonian vector fields exhaust the tangent bundle. In that case, the Poisson bracket comes from a symplectic form $\om$ on $M$ in the usual way  \cite{AM}. 

Suppose a Lie algebra  $\g$ acts on a Poisson manifold $M$ in strongly  Hamiltonian fashion. 
This means that there exist  Lie algebra homomorphisms  $X\mapsto X^M$ from  $\g$  to the space $\Gm(M,TM)$ of vector fields on $M$ and  $X\mapsto J_X$ from $\g$ to 
 $\cin(M)$,  with the property 
$X^M=\xi_{J_X}$. 
The functions $J_X$ may  be assembled into a so-called  momentum map $J:S\raw \g^*$,
defined by $\la J(\sg),X\ra=J_X(\sg)$. Here  $\g^*$ is the dual vector space of the Lie algebra $\g$. This $\g^*$ is canonically a Poisson manifold under the Lie--Poisson
bracket, defined on linear functions (hence elements of $\g^{**}=\g$) by 
the Lie bracket.
It follows  that $J$ is a  Poisson map. Note that a smooth map between two Poisson manifolds is called  Poisson
when its  pullback is a Lie algebra homomorphism (and anti-Poisson when it is an
anti homomorphism). 
It may happen that  the $\g$ action comes from a $G$ action
$G\circlearrowright M$,  where $G$ is a Lie group with Lie algebra $\g$: in that case,  one has $X^Mf(\sg)=df(\exp(-tX)\sg)/dt|t=0$. The $G$ action is called
strongly  Hamiltonian whenever the associated $\g$ action is. 

We now specialize to the case where $M$ is symplectic.   The symplectic quotient or reduced space defined by the $G$ action, or physically by the constraint
$J=0$, is $M^0=J\inv(0)/G$. In case that $0$ is a regular value of $J$ and 
the $G$ action is proper and free on $J\inv(0)$,
$M^0$ is a manifold,  which moreover carries a unique symplectic form $\om^0$ with the property
$i^*\om=\pi^*\om^0$. Here $i:J\inv(0) \hookrightarrow  M$ is the inclusion and
$\pi: J\inv(0)\raw M^0$ is the projection map. Thus Marsden--Weinstein reduction
produces a new symplectic manifold $(M^0,\om^0)$ from a given symplectic manifold
$(M,\om)$  equipped with a strongly  Hamiltonian  $G$ action \cite{AM,MW,MW2,Mey}.
If the stated assumptions are not met,  singularities may arise in the reduced space (cf.\ \cite{Ober,Pfl,SL}). 
 \section{Symplectic  dual pairs as arrows}\label{Sdp}
On the classical side, a bimodule over a pair $P, Q$ of Poisson manifolds is
by definition a  so-called  symplectic 
dual  pair  \cite{K1,W1}   $Q\law M \raw P$, simply called a dual pair in what follows.
Here $M$ is a symplectic manifold,
 the map $Q\law M$ is 
Poisson,  and $M \raw P$ is anti-Poisson. 
 Furthermore, the pullback of any function on $P$
should Poisson-commute on $M$ with  the pullback of any function on $Q$.
One of the motivating examples of a dual pair is $G\backslash
 M\law M\stackrel{J}{\raw}\g^*_-$,
 obtained from a strongly Hamiltonian $G$ action on $M$ with momentum map 
$J$.\footnote{In general, we write $P^-$ for a Poisson manifold $P$ equipped with minus
its Poisson bracket, but we write $\g^*_-$ for $(\g^*)^-$.} 
Similarly, $\g^*_-\law M^-\raw G\backslash M$ is a dual pair. 

  Two $Q$-$P$ dual pairs
$Q\stackrel{q_i}{\law}\til{M}_i\stackrel{p_i}{\raw}P$, $i=1,2$, are said to be 
isomorphic when there is a symplectomorphism $\phv:\til{M}_1\raw
\til{M}_2$ for which $q_2\phv=q_1$ and $p_2\phv=p_1$.
We now interpret the equivalence class of a dual pair $Q\law M \raw P$ as an arrow
from $Q$ to $P$. Two compatible dual pairs $Q\law M_1\raw P$ and
$P\law M_2\raw R$ can be composed when firstly
 $M_1\x_P M_2$  is a coisotropic submanifold of $M_1\x M_2$, and secondly the
associated symplectic quotient of $M_1\x_P M_2$ by its canonical foliation is
a manifold. We then denote the product of the dual pairs in question by 
  $P\law M_1\otc_{P} M_2\raw R$. This product is well defined on equivalence classes,
where it is associative, since the operation $\otc$ is associative up to isomorphism.
For example, when $G$ is connected the product of the dual pairs $G\backslash M\law M\raw\g^*_-$ and $\g^*_-\hookleftarrow  0\raw pt$ is 
$G\backslash M\law M^0\raw pt$,
where $M^0$ is the Marsden--Weinstein quotient $J\inv(0)/G$ as before. 

As explained in \cite{OBWF} (see also \cite{BRW}), one can impose certain regularity conditions
on both Poisson manifolds and dual pairs, which guarantee that all products exist
and that one has identity arrows from $P$ to $P$. Thus one obtains a category
\textsf{Poisson} 
whose objects are (regular) Poisson manifolds and whose arrows are equivalence
classes of (regular) dual pairs. The regularity condition on Poisson manifolds
is very mild, and it is actually quite hard to construct an example that fails to satisfy it. 
On the other hand, many dual pairs one would like to use are not regular, such as
$pt\law M\raw pt$, where $pt$ is the space consisting of a point.
Also, although  $G\backslash M\law M\raw\g^*_-$ is regular, $pt\law M\raw \g^*_-$ is not. 
Nonetheless, the product of $pt\law M\raw\g^*$ and $\g^*\hookleftarrow 0\raw pt$
is well defined, and equal to 
\begin{equation}
(pt\law M\raw\g^*_-)\otc_{\g^*_-}(\g^*_-\hookleftarrow 0\raw pt)\cong 
pt\law M^0\raw pt. \label{comp}
\end{equation}

 Another example
is the dual pair $X\stackrel{\pi}{\law} T^*X\stackrel{f\circ\pi}{\raw} Y$ defined by a smooth map $X\stackrel{f}{\raw} Y$.
Here $X$ and $Y$ are manifolds with zero Poisson bracket,  $f$ is smooth, and
$T^*X$ has the canonical Poisson structure.  
The product of $X\law T^*X\raw Y$ with the dual pair
$Y\law T^*Y\raw Z$ induced by $Y\stackrel{g}{\raw} Z$ is 
\begin{equation}
(X\law T^*X\raw Y) \otc_Y (Y\law T^*Y\raw Z)\cong 
X\law T^*X\stackrel{g\circ f\circ\pi}{\longrightarrow} Z.
\end{equation}

Note that the dual pairs defined by a $G$ action $G\car M$ and by a map
$X\stackrel{f}{\raw} Y$ are both special cases of a very general functorial construction
involving Lie groupoids \cite{NPLLMP}. 
Such  examples indicate that
products of dual pairs lying in a certain class 
often make sense when the regularity condition is not satisfied.
Thus in the present paper  we shall not  impose the
regularity conditions on dual pairs,  refraining from a complete categorical
structure. It will still be possible to map arrows of the above type into arrows in the category
\textsf{KK} defined below, and to check functoriality of this map, interpreted as quantization,
with respect to the product $\otc$. It is in this rather pragmatic sense that 
the notion of functoriality will be understood in what follows.
 \section{The Guillemin--Sternberg conjecture}
Guillemin and Sternberg \cite{GS82}
considered the case in which
the symplectic manifold  $M$ is compact,  prequantizable, and equipped with a positive-definite complex polarization $\mathcal{J}$. Recall that a symplectic
manifold $(M,\om)$ is called prequantizable when the cohomology class
$[\om]/2\pi$ in $H^2(M,\R)$ is  integral, i.e., lies in the image of $H^2(M,\Z)$
under the natural homomorphism $H^2(M,\Z)\raw H^2(M,\R)$. In that case, there exists
a line bundle $L_{\om}$ over $M$ whose first Chern class $c_1(L_{\om})$ maps to $[\om]/2\pi$
under this  homomorphism; $L_{\om}$ is called the prequantization line bundle over $M$.
In general, this bundle is not unique. 

Under these circumstances,  the quantization  operation $Q$ is well-defined through geometric quantization \cite{GGK}: one picks a connection $\nabla$ on $L_{\om}$ whose curvature is $\om$,
and defines  the Hilbert space $H(M)$ as  the space $H=H^0(M,L_{\om})$ of polarized  sections of   $L_{\om}$  (i.e., of sections annihilated by all $\nabla_X$, $X\in \mathcal{J}$).

 Now suppose that $M$ carries a strongly Hamiltonian
action $G\car M$ of a compact Lie group $G$ that leaves $\mathcal{J}$ invariant. 
The Hilbert space $H(M)$ then   carries a natural unitary \rep\ of $G$ determined  by the classical data, as polarized sections of $L_{\om}$ are mapped into each other by the pullback of the $G$ action.
 Moreover, it turns out that the reduced space $M^0$ inherits all relevant structures on $M$ (except, of course, the $G$ action), so that it is quantizable as well, in the same fashion. Thus (\ref{QR0}) becomes, in obvious notation,  $H^0(M,L_{\om})^G\cong H^0(M^0,L_{\om}^0)$, which  Guillemin and Sternberg indeed managed to prove. The  idea of the proof is to define a map from $H^0(M,L_{\om})^G$ to $H^0(M^0,L_{\om}^0)$
by simply restricting a $G$ invariant polarized section of $L_{\om}$ to $J\inv(0)$; this map is then
shown to be an isomorphism \cite{GS82}. 

\section{$\spinc$ structures and Dirac operators}
The new approach to geometric quantization 
mentioned in the Introduction is based on the notion of a $\spinc$ structure
on  $M$, which we briefly recall.\footnote{Such a structure may more generally be defined on a real vector bundle $E$ 
over $M$;  when  $E$ is the tangent bundle $TM$ we obtain  the special case discussed in the main text.} 
A large number of approaches to  $\spinc$ structures exist, of which the ones relating
this concept to K-theory \cite{ABS,LM}, to K-homology \cite{BD,HR}, to KK-theory
\cite{CS}, to E-theory \cite{Con} (all these approaches are, in turn, closely linked to
index theory),  and to Morita equivalence of
$C^*$-algebras \cite{GVH,Ply}  are particularly relevant to our theme. 
We will return to some of these in due course, but for the moment
 a purely  differential-geometric approach  is appropriate \cite{Dui,GGK}. 

Firstly, the compact Lie group $\spinc(n)$ is a nontrivial central extension of $SO(n)$ by $U(1)$, defined as $\spinc(n)=\mathrm{Spin}(n)\x_{Z_2} U(1)$, where $\mathrm{Spin}(n)$ is the usual twofold cover of $SO(n)$, and  $Z_2$ is seen as the subgroup $\{(1,1),(-1,-1)\}$
of $\mathrm{Spin}(n)\x U(1)$. Thus one has the obvious homomorphisms
$\pi:\spinc(n)\raw SO(n)\cong \mathrm{Spin}(n)/Z_2$, 
given by projection on the first factor, and $\det: \spinc(n)\raw U(1)$, defined by
$[x,z]\mapsto z^2$. 

Let  $n=\dim(M)$. A $\spinc$ structure $(P,\cong)$ on $M$ is by definition a
principal $\spinc(n)$-bundle $P$ over $M$  with an isomorphism
 $P\x_{\pi} \R^n\cong TM$ of vector bundles. Here  the bundle on the left-hand side
is the bundle associated to $P$ by the defining \rep\ of $SO(n)$.
 Various structures on $M$ canonically induce a $\spinc$ structure on $M$, such as a Spin structure or an almost complex structure.  Note that a $\spinc$ structure on $M$, when it exists, is not unique: up to homotopy, the class of possible  $\spinc$ structures on $M$ 
(with given orientation) is parametrized by the Picard group $H^2(M,\Z)$ \cite{GGK}. 

A $\spinc$ structure defines a number of vector bundles over $M$ associated to $P$ by
various \rep s of $\spinc(n)$. The first of these, which is isomorphic to the bundle
$TM$,  has just been mentioned.\footnote{It   induces both an orientation and a 
Riemannian metric on $M$, by transferring the standard orientation and  metric on $\R^n$ to $E$.  Conversely, given an orientation and a   Riemannian metric on $M$, one should require a $\spinc$ structure on $M$ to be compatible with these.} The second 
is the  canonical line bundle $L=P\x_{\det} \C$   associated to $P$ by the defining \rep\ of $U(1)$. Thirdly, $\spinc(n)$ has a canonical unitary \rep\ $\Delta_n$ on 
a finite-dimensional Hilbert space $S$, the so-called (complex)
spin \rep, which for odd $n$ is irreducible, and for even $n$ decomposes into two
irreducibles $\Dl_n=\Dl_n^+ \oplus \Dl_n^-$ on $S=S^+\oplus S^-$. Thus one has an associated spinor bundle
$\CS=P\x_{\Dl_n}S$, which  for even $n$  decomposes into the direct sum 
$\CS^{\pm}=P\x_{\Dl_n}S^{\pm}$.  
Thus the physical interpretation of $\spinc$ structures involves  gravity, electromagnetism, and fermions.

A  $\spinc$ structure on $M$ defines a vector bundle action $TM \raw \mathrm{End}(\CS)$
by Clifford multiplication, since both $TM$ and $\CS$ are subspaces of the Clifford
bundle $Cl(TM)$ over $M$. This action may be seen as a map $c:\Gm(TM\ot_M\CS)\raw
\Gm(\CS)$. Furthermore, a connection on $P$ induces one on $\CS$,
which amounts to  a covariant derivative
$\nabla: \Gm(\CS)\raw \Gm(T^*M\ot_M\CS)$. Identifying $T^*M$ with $TM$ through the Riemannian metric $g$ determined by the $\spinc$ structure and composing these maps
yields the Dirac operator
$$
\DS: \Gm(\CS)\stackrel{\nabla}{\raw} \Gm(T^*M\ot_M\CS)\stackrel{g\ot\mathrm{id}}{\raw}
\Gm(TM\ot_M\CS)\stackrel{c}{\raw} \Gm(\CS).
$$

This elliptic first-order linear differential operator  is formally self-adjoint, and can be turned into a bounded self-adjoint operator
$\tilde{\DS}=\DS/\sqrt{1+\DS^*\DS}:L^2(\CS)\raw L^2(\CS)$, where $L^2(\CS)$
stands for the Hilbert space of $L^2$-sections of the vector bundle $\CS$. 
When $M$ is even-dimensional, $\DS$ is odd with repsect to the decomposition
$\CS=\CS^+ \oplus\CS^-$, so that one obtains the chiral Dirac operator
$\DS^+:\Gm(\CS^+)\raw \Gm(\CS^-)$, with formal adjoint  $\DS^-:\Gm(\CS^-)\raw \Gm(\CS^+)$, by restriction. Similarly, one has  $\til{\DS}\mbox{}^{\pm}:L^2(\CS^{\pm})\raw L^2(\CS^{\mp})$.
\section{Bott's definition of quantization}
The first step in Bott's definition of quantization is to canonically associate a $\spinc$ structure to a given
symplectic and prequantizable manifold  $(M,\om)$ \cite{GGK,Mei}. First, one picks an almost complex structure $\CJ$ on $M$ that is compatible with $\om$ (in that $\om(-,\CJ-)$ is positive definite and symmetric, i.e., a metric). This $J$ canonically induces a $\spinc$ structure $P_{\CJ}$ on $TM$ \cite{Dui,GGK}, but this is not the right one to use here. 
The $\spinc$ structure $P$ needed to quantize $M$ is the one obtained by twisting $P_{\CJ}$ with the prequantization line bundle $L_{\om}$. This means (cf.\ \cite{GGK}, App.\ D.2.7) that $P=P_{\CJ}\times_{\ker(\pi)} U(L_{\om})$,
where $\pi: \spinc(n)\raw SO(n)$ was defined in the preceding section 
(note that $\ker(\pi)\cong U(1)$), and $U(L_{\om})\subset L_{\om}$ is the unit circle bundle.\footnote{In fact, this construction needs to be corrected in some cases
\cite{GGK,Par2}, but this correction complicates the statement of the Guillemin--Sternberg conjecture, and will not be discussed here.} 

When $M$ is compact, the operators $\til{\DS}\mbox{}^{\pm}$ determined by the $\spinc$ structure $(P,\cong)$
have finite-dimensional  kernels, whose dimensions  define the quantization of $(M,\om)$ as
\begin{equation}
Q(M,\om)= \mathrm{index}(\DS^+)=\dim\ker (\DS^+)-\dim\ker (\DS^-).\label{defQ}
\end{equation}

In fact, the corresponding Hilbert space operators  $\til{\DS}\mbox{}^{\pm}$ are Fredholm, 
and by elliptic regularity $\mathrm{index}(\DS^+)$ coincides with the  Fredholm index 
$\dim\ker (\til{\DS}\mbox{}^{+})-\dim\ker (\til{\DS}\mbox{}^{-})$ of $\til{\DS}\mbox{}^{+}$. 
  This notion of quantization just associates
an integer to $(M,\om)$. This number turns out to be independent of the choice of the
$\spinc$ structure on $M$, as long as it satisfies the above requirement, and  is
entirely determined by the cohomology class $[\om]$ (as remarked earlier, this is not true for 
the $\spinc$ structure and the  associated Dirac operator itself) \cite{GGK}.

This definition of quantization gains in substance when a compact Lie group $G$ acts
on $M$ in strongly  Hamiltonian fashion. In that case, the pertinent $\spinc$ structure
may be chosen to be $G$ invariant, and the  spaces $\ker (\DS^{\pm})$ 
are finite-dimensional complex $G$ modules. 
 Hence 
\begin{equation}
\Gindex(\DS^+)=[\ker (\DS^+)]-[\ker (\DS^-)] \label{Gindex}
\end{equation}
defines an element of the \rep\ ring $R(G)$ of $G$.\footnote{$R(G)$ is defined as the abelian group with one generator $[L]$ for each
finite-dimensional complex \rep\ $L$ of $G$, and relations $[L]=[M]$ when
$L$ and $M$ are equivalent and $[L]+[M]=[L\oplus M]$. The tensor product
of \rep s defines a ring structure on $R(G)$.} Thus, the quantization of
$(M,\om)$ with associated $G$ action may be defined as
\begin{equation}
Q(G\car M,\om)= \Gindex(\DS^+)\in R(G). \label{defGQ}
\end{equation}
As before, this element only depends on $[\om]$ (and on the $G$ action). 
The same definition arises from the Hilbert space setting: the
Hilbert spaces $L^2(\CS^{\pm})$ carry  unitary \rep s $U^{\pm}$  of $G$ in the obvious way, and the bounded Dirac operators $\til{\DS}\mbox{}^{\pm}$  are equivariant under
these, so that $\ker (\til{\DS}\mbox{}^{\pm})$ are unitary $G$ modules. 
Replacing  $\DS^{\pm}$ in (\ref{Gindex})  by $\til{\DS}\mbox{}^{\pm}$ then yields an element
of the ring of unitary finite-dimensional \rep s of $G$, which for a compact group
is the same as $R(G)$. 
 When $G$ is trivial, one may identify $R(e)$ with $\Z$ through the map
$[V]-[W]\mapsto \dim(V)-\dim(W)$, so that  (\ref{defQ})  emerges as a special case of (\ref{defGQ}).

In this setting, the Guillemin--Sternberg conjecture makes sense  
 as long as $M$ and $G$ are compact. The Hilbert space
$H^0(M,L_{\om})^G$ in the original version of the conjecture is now replaced by the image
$Q(G\car M,\om)_0$
of $Q(G\car M,\om)$ in $\Z$ under the map $[V]-[W]\mapsto \dim(V_0)-\dim(W_0)$, where
$V_0$ is the $G$ invariant part of $V$, etc. The right-hand side of the conjecture
is the quantization of  the reduced space $M^0$ (which inherits a $\spinc$ structure
from $M$) according to (\ref{defQ}). Denoting the pertinent Dirac operator on $M^0$
by $\DS_0$, the Guillemin--Sternberg conjecture in the setting of Bott's definiton of
quantization is therefore simply
\begin{equation}
\Gindex(\DS^+)_0= \mathrm{index}(\DS^+_0). \label{G0}
\end{equation}
In this form, the conjecture was proved in  \cite{Mei}; it even holds
 when 0 fails to be a regular value of $J$ \cite{MS}. Also see \cite{GGK,Par1} for other proofs and 
further references.

Bott's definition of quantization (\ref{defGQ}) or (\ref{defQ}) isn't actually all that far removed from the traditional idea of associating a group \rep\ on a Hilbert space with a strongly
Hamiltonian action on a symplectic manifold. In fact, when the symplectic form $\om$
is sufficiently large, the space $\ker (\DS^-)$ tends to vanish \cite{Brav1}, so that 
$Q(G\car M,\om)$ is really a \rep\ of $G$, up to  isomorphism. This is relevant in the 
semiclassical regime, where one quantizes $(M,\om/\hbar)$ for small values of $\hbar$.
\section{From quantization to KK-theory}
To motivate the use of Kasparov's bivariant K-theory, or KK-theory,  in the light of 
the Guillemin--Sternberg conjecture and Bott's definition of quantization, let us
recall a result  from functional analysis (see, e.g., \cite{Dou}). 
Recall that a bounded operator $F:H^+\raw H^-$ between two Hilbert spaces
is called Fredholm when it is invertible up to compact operators, that is, when there
exists a bounded operator $F':H^-\raw H^+$, called a parametrix of $F$, such that $FF'-1$ and $F'F-1$ are compact
operators on $H^-$ and $H^+$, respectively. A key result is then that the 
space $\mathcal{F}(H^+,H^-)/\stackrel{h}{\sim}$ 
of homotopy equivalence classes $[F]$ of Fredholm operators $F$ (where the notion of homotopy is defined with respect to operator-norm continuous paths in the space of all Fredholm operators) is homeomorphic to $\Z$, where the pertinent homeomorphism
is given by $[F]\mapsto\mathrm{index}(F)$.  

Hence in Bott's definition of quantization (\ref{defQ}) we may work with
$[\til{\DS}\mbox{}^{+}]$ instead of with $\mathrm{index}(\DS^+)$ ($=\mathrm{index}(\til{\DS}\mbox{}^{+})$). Thus we put
\begin{equation}
Q(pt\law M\raw pt)= [\til{\DS}\mbox{}^{+}] . 
\label{Bottpp}
\end{equation}
 
As indicated by the notation, we  regard the right-hand side of (\ref{Bottpp}) as the quantization of
(the isomorphism class of) the dual pair on the left-hand side. It will become clear shortly that
this homotopy class is an element of the Kasparov group $KK(\C,\C)$, where we regard $\C$ as the \ca\ that quantizes the Poisson manifold $pt$. This group is isomorphic to $\Z$, and
the image of $[F]$\footnote{{\samepage More precisely, of the homotopy class $[F,H^+,H^-]$,
where $H^{\pm}$ are $\C$-$\C$ Hilbert bimodules under the action $z\mapsto z1$,
$z\in \C$.}}
 under  the isomorphism $KK(\C,\C)\raw\Z$ is precisely
$\mathrm{index}(F)$.  
Clearly, this  isomorphism  links  (\ref{Bottpp}) to (\ref{defQ}). 

To generalize this idea to more complicated dual pairs, we need Kasparov's theory \cite{Kas1} (see also \cite{Bla} for a full treatment and
\cite{Con,Hig,Ska} for  very useful introductions), which is a systematic machinery for dealing with homotopy classes of generalized Fredholm operators. The first step is to generalize 
the notion of a Hilbert space, which we here regard as a Hilbert $\C$-$\C$ bimodule, to 
the concept of a Hilbert $A$-$B$ bimodule, where $A$ and $B$ are
separable  \ca s (which in our setting emerge  as the quantizations of Poisson manifolds $P$ and $Q$).
The correct  generalization  was introduced by Rieffel in a different context \cite{Rie}, and has already been used in the theory of constrained quantization in \cite{NPL}. 

  An $A$-$B$ Hilbert bimodule is an algebraic $A$-$B$
bimodule $E$ (where $A$ and $B$ are seen as complex algebras, so that
$E$ is a complex linear space) with a compatible $B$-valued inner
product. This is a sesquilinear map $\langle \,,\,\ra :E\x E\raw
B$, linear in the second and antilinear in the first entry, satisfying
$\langle x,y \ra^* =
\langle y, x\ra$, $\langle x, x \ra \geq 0$,
and $\langle x,x\ra = 0$ iff $x=0$.  The compatibility of the
inner product with the remaining structures means that firstly $E$ has
to be complete in the norm $\| x\|^2= \|\langle x,x\ra \|$, secondly
that $\langle x,yb\ra =\langle x,y\ra  b$, and thirdly that
$\langle a^* x,y\ra=\langle x,ay\ra$ for all $x,y\in E$, $b\in B$, and $a\in A$.
The latter condition may be expressed by saying  that $a$ is adjointable, with adjoint $a^*$; 
this is a nontrivial condition even when $a$ is bounded (note that an adjointable
operator is automatically bounded). The best example of all this is
the  $A$-$A$ Hilbert bimodule $E=A$,  with the obvious actions and the inner product
$\langle a,b\ra=a^*b$.

An $A$-$\C$  Hilbert bimodule is simply a Hilbert space equipped with a \rep\ of $A$.
A $\C$-$B$ Hilbert bimodule is called a Hilbert $B$ module, or Hilbert $C^*$-module
over $B$.

Adjointable operators  on an $A$-$B$ Hilbert bimodule $E$ are the analogues of bounded operators on a Hilbert space; the collection of all adjointable operators indeed forms
a \ca.  The role of compact operators on $E$ is played by  operators that can be approximated in norm by linear combinations of rank one operators of the form $z\mapsto x \langle y,z\ra$ for $x,y\in E$ (such operators are automatically
adjointable). Again, as for Hilbert spaces, the space of all compact operators on $E$ is a \ca.
In the example ending the preceding paragraph, the left $A$ action turns out to be by
compact operators.  A Fredholm operator, then, is  an adjointable  operator that is 
invertible up to compact operators. 

Now an  $A$-$B$  Kasparov bimodule is a pair of  
countably generated  $A$-$B$ Hilbert bimodules $(E^+,E^-)$  with  an  `almost'  Fredholm operator $F:E^+\raw E^-$ that `almost' intertwines 
the $A$ actions on $E^+$ and $E^-$. The first condition  means that  there is
an adjointable operator  $F':H^-\raw H^+$  such that $a(FF'-1)$ and $a(F'F-1)$ are compact  for all $a\in A$,   and the second states that $aF-Fa$ is  compact  for all $a\in A$. 
With the structure of $E^{\pm}$ as $A$-$B$ Hilbert bimodules understood, we denote such a Kasparov bimodule simply by $(F,E^+,E^-)$. 

 For $B=\C$ this is sometimes called a Fredholm module \cite{Con}. 
A key  example of a Fredhom module is given by $E^{\pm}=L^2(\CS^{\pm})$,
and $F=\til{\DS}\mbox{}^{+}$.
When $M$ is compact, this works for  both $A=\C$ and $A=C(M)$,
 but  when $M$ isn't  one must take $A=C_0(M)$. For general $A$ and $B$, it follows
from the definitions that if $A$ acts on $E$ by compact operators, 
then the choice $F=0$ 
yields a  Kasparov bimodule. This applies, for instance, to the  $A$-$A$ Hilbert bimodule $(E^+=A, E^-=0)$.

A  homotopy of  $A$-$B$  Kasparov bimodules is an $A$-$C([0,1],B)$ Kasparov bimodule.
The ensuing set $KK(A,B)$ of homotopy classes of $A$-$B$  Kasparov bimodules
may  more conveniently be described as the quotient of the set of all 
$A$-$B$  Kasparov bimodules by the equivalence relation
generated by  unitary equivalence,  translation of $F$ along norm-continuous paths
(of almost intertwining almost Fredholm operators),
and the addition of degenerate Kasparov bimodules. The latter are those
for which the operators $aF-Fa$, $a(FF'-1)$ and $a(F'F-1)$ are 
not merely compact but zero  for all 
$a\in A$.  Using the polar decomposition, one may 
 always choose  representatives for which all $(F'-F^*)a$ are  compact
(so that $F$ is almost unitary), and this is often
included in the definition of a Kasparov bimodules. In that case, the condition
that $(F'-F^*)a=0$ is added to the definition of a degenerate Kasparov bimodule. 

It is not difficult to see that $KK(A,B)$ is an abelian group; the group operation
is the direct sum of both bimodules and operators $F$, and the inverse of
the class  of a Kasparov bimodule is found by 
swapping $E^+$ and $E^-$ and replacing  $F:E^+\raw E^-$ by
its parametrix $F':E^-\raw E^+$. 
Moreover, with respect to \sthom s between \ca s 
the association $(A,B) \mapsto KK(A,B)$ is contravariant
in the first entry, and covariant in the second. 

Let us note that  for any \ca\ $A$ the group $KK(\C,A)$
is naturally isomorphic to the algebraic K-theory group
$K_0(A)$.\footnote{When $A$ has a unit, $K_0(A)$
may be defined as the abelian group with one generator $[E]$ for each
finitely generated projective (f.g.p.)  right module over $A$, and relations
$[E]=[E']$ when $E$ and $E'$ are isomorphic, and $[E]+[E']=[E\oplus E']$.
For example, when $X$ is a compact Hausdorff space one has
$K_0(C(X))=K^0(X)$, the topological K-theory of Atiyah and Hirzebruch
\cite{Kar}. When $A$ has no unit, $K_0(A)$ is defined as the kernel
of the canonical map $K_0(\til{A})\raw K_0(\C)$, where $\til{A}=A\oplus\C$
is the unitization of $A$.}  Hence as far as $K_0$ is concerned,  K-theory is a special case of KK-theory. Explicitly, the isomorphism $KK(\C,A)\raw K_0(A)$
is the generalized index map\footnote{The representatives $F$ and $F'$ of their respective  homotopy classes have to be chosen such that their kernels in 
the $A$ modules $E^-$ and $E^+$ are indeed f.g.p.}
\begin{equation}
[F,E^+,E^-]\mapsto [\ker(F)]-[\ker(F')].  \label{KKiso}
\end{equation}

A remarkable aspect of Kasparov's theory is the existence of a  product
$$KK(A,B)\x KK(B,C)\raw KK(A,C),$$
which is functorial in all conceivable ways. Disregarding $F$, this 
would be  easy to define, since one feature of algebraic
bimodules that survives in the Hilbert case is the existence of a
bimodule tensor product \cite{Rie}: from an $A$-$B$ Hilbert
bimodule $E$ and a $B$-$C$ Hilbert bimodule $\til{E}$ one can form an
$A$-$C$ Hilbert bimodule $E\hat{\ot}_B \til{E}$, called the interior tensor
product of $E$ and $\til{E}$. 
However, the composition of the almost Fredholm operators in question is
too complicated to be explained here (see \cite{Bla,Con,CS,Hig,Kus,Ska}).
 In any case,  this product leads to the category \textsf{KK},
whose objects are separable \ca s, and whose arrows are 
Kasparov's  KK-groups.

To close this section, let us mention that we only use the `even' part of KK-theory;
in general, each KK group  is $\Z_2$ graded, and what we have called
$KK(A,B)$ is really $KK_0(A,B)$. This restriction is possible  because symplectic manifolds 
happen to be  even-dimensional.
 \section{The Guillemin--Sternberg conjecture revisited}
Let us return to a strongly Hamiltonian group action $G\car M$, with associated
dual pair $pt\law M\raw \g^*_-$. To quantize this dual pair, we first note that the
quantization of the Poisson manifold $\g^*$ is the group \ca\ $C^*(G)$ 
\cite{Rie3,NPL}; this is probably the best understood example in \ca ic quantization 
theory.\footnote{Here $C^*(G)$ is a suitable completion of the convolution algebra on $G$ determined by a Haar measure \cite{Dix,NPL}.}
 Although this holds for any $G$ with given Lie algebra, to obtain a unique functor we assume $G$ to be connected and simply connected. 
 Hence the quantization of the dual pair $pt\law M\raw \g^*_-$ should be an element of
the Kasparov group $KK(\C,C^*(G))\cong K_0(C^*(G))$.

 When $G$ is compact, which we assume throughout the remainder of this section, 
one may identify $K_0(C^*(G))$ with the \rep\ ring $R(G)$;
this is because finitely generated projective modules over $C^*(G)$ may be identified
with finite-dimensional unitary \rep s of $G$. 
Now assume that $M$ is compact as well. Seen as an element of $R(G)$, the 
quantization of $pt\law M\raw \g^*_-$ is given by $\Gindex(\DS^+)$,
as in (\ref{Gindex}); this is just a reinterpretation of Bott's definition (\ref{defGQ})
of quantization. It is slightly more involved to explain  the 
quantization of $pt\law M\raw \g^*_-$ when it is  seen as an element of $KK(\C,C^*(G))$. Firstly, one turns the Hilbert  spaces
$L^2(\CS^{\pm})$ into Hilbert $C^*(G)$ modules, as follows \cite{BCH,Val,Val03}.

The canonical $G$ actions $U^{\pm}$ on $L^2(\CS^{\pm})$ induce
right actions $\pi^{\pm}$ of $C^*(G)$ by $$\pi_-^{\pm}(f)=\int_G dx\, f(x) U^{\pm}(x\inv),$$
where $f\in C(G)$ (the action of a general element of $C^*(G)$ is then defined by
continuity). Furthermore, one obtains a $C^*(G)$ valued inner product on
$L^2(\CS^{\pm})$ by the formula 
\begin{equation}
\la\ps,\phv\ra: x\mapsto (\ps, U^{\pm}(x)\phv), \label{Gip}
\end{equation}
which defines an element of $C(G)\subset C^*(G)$. 
Completing $L^2(\CS^{\pm})$ in the norm 
\begin{equation}
\| \ps\|^2=\| \la\ps,\ps\ra\|_{C^*(G)} \label{norm}
\end{equation}
then yields Hilbert $C^*(G)$ modules $E^{\pm}(\CS)$. The operator
$\til{\DS}\mbox{}^{+}:L^2(\CS^+)\raw L^2(\CS^-)$ extends to an adjointable operator
$\hat{\DS}\mbox{}^+:E^+(\CS)\raw E^-(\CS)$ by continuity, and the triple $(\hat{\DS}\mbox{}^+,E^+(\CS),E^-(\CS))$
defines a $\C$-$C^*(G)$ Kasparov bimodule, whose homotopy class is the
desired element of $KK(\C,C^*(G))$, i.e.,
\begin{equation}
Q(pt\law M\raw \g^*_-)= [\hat{\DS}\mbox{}^+,E^+(\CS),E^-(\CS)]. \label{Q1}
\end{equation}
The canonical isomorphism 
$KK(\C,C^*(G))\raw K_0(C^*(G))=R(G)$ given by (\ref{KKiso})
indeed maps this element to $\Gindex(\DS^+)$.

Apart from the dual pair $pt\law M\raw \g^*_-$, the momentum map associated to
the action $G\car M$ equally well leads to a dual pair $\g^*_-\law M^-\raw pt$. 
This is to be quantized by an element of $KK(C^*(G),\C)\cong K^0(C^*(G))$,
the so-called Kasparov \rep\ ring of $G$ (cf.\ \cite{HR}). 
This time, we interpret the Hilbert spaces $L^2(\CS^{\pm})$ as 
$C^*(G)$-$\C$ Hibert bimodules, where the pertinent \rep s $\pi^{\pm}$ of $C^*(G)$
are given by a very slight adaptation of the procedure sketched in the preceding
paragraph: to obtain left actions instead of right actions, we now put
$\pi^{\pm}(f)=\int_G dx\, f(x) U^{\pm}(x)$. Since $\til{\DS}\mbox{}^{+}U^+(x) =U^-(x) \til{\DS}\mbox{}^{+}$
for all $x\in G$,  one now has  $\til{\DS}\mbox{}^{+}\pi^+(f) =\pi^-(f) \til{\DS}\mbox{}^{+}$ for all $f\in C^*(G)$.
Since $\til{\DS}\mbox{}^{+}$ is Fredholm one thus obtains an element 
$[\til{\DS}\mbox{}^{+}, L^2(\CS^+), L^2(\CS^-)]$ of  $KK(C^*(G),\C)$, which we regard as
the quantization of the dual pair $\g^*_-\law S^-\raw pt$. 

The very simplest example is the dual pair $\g^*_-\hookleftarrow 0\raw pt$, whose
quantization is just 
\begin{equation}
Q(g^*_-\hookleftarrow 0\raw pt)=[0,\C,\C], \label{Q2}
\end{equation}
 where the $C^*(G)$-$\C$ Hilbert bimodules
$\C$ carry the trivial \rep\ of $G$. A simple computation of the Kasparov product
$$KK(\C,C^*(G))\x KK(C^*(G)),\C)\raw KK(\C,\C)\cong K_0(\C)\cong\Z$$
 yields
\begin{equation}
[\hat{\DS}\mbox{}^+,E^+(\CS),E^-(\CS)]\x [0,\C,\C]= \Gindex(\DS^+)_0, \label{Q3}
\end{equation}
cf.\ (\ref{G0}) and preceding text.
In fact, $y \x [0,\C,\C]$ is just the image of $y$ 
under the map $KK(\C,C^*(G))\raw KK(\C,\C)$ functorially induced by
the $\mbox{}^*$-homomorphism $C^*(G)\raw \C$ given by the trivial \rep\ of $G$.

As explained around (\ref{Bottpp}), 
if we identify $KK(\C,\C)$ with $\Z$ as above, the reduced space $M^0$
is quantized  by 
\begin{equation}
Q(pt\law M^0\raw pt)= \mathrm{index}(\DS^+_0).  \label{Q01}
\end{equation}
Combining (\ref{comp}), (\ref{Q1}), (\ref{Q2}), (\ref{Q3}), and  (\ref{Q01}),
we see that  the functoriality condition
\begin{eqnarray}
 Q(pt\law M\raw \g^*_-)\x Q(g^*_-\hookleftarrow 0\raw pt) & = & \nn \\ 
Q((pt\law M\raw\g^*_-)\otc_{\g^*_-}(\g^*_-\hookleftarrow 0\raw pt))
& &  \label{GSrev}
\end{eqnarray}
is precisely the  Guillemin--Sternberg conjecture (\ref{G0}). 
\section{Guillemin--Sternberg for noncompact groups}
The above  reformulation of the  Guillemin--Sternberg conjecture as a special case of the functoriality of  Bott's definiton of quantization paves the way for far-reaching
generalizations of this conjecture. Firstly, one can now consider noncompact
$G$ and $M$, as long as the $G$ action on $M$ is proper. It is convenient to
use the language of K-homology (cf.\ \cite{HR}). The K-homology group
of a manifold $M$ is just defined as the Kasparov group $K_0(M)=KK(C_0(M),\C)$.
A $\spinc$ structure on $M$ defines an element  $[\til{\DS}\mbox{}^{+}]$
of $K_0(M)$ through its associated Dirac operator. This so-called fundamental class never vanishes. It
is independent of the connection picked to define $\DS$, and is the analogue
in K-homology of the fundamental class in ordinary homology defined by the orientation of $M$ \cite{HR}. From this point of view, 
Bott's  quantization (\ref{defQ}) of $(M,\om)$, which in our setting is the quantization
of the dual pair $pt\law M\raw pt$, is the image of the fundamental class
of $M$ determined by the symplectic structure as explained, under the map $KK(C_0(M),\C)\raw KK(\C,\C)$ obtained by forgetting the
$C_0(M)$ actions on $L^2(\CS^{\pm})$ (followed by the isomorphism $KK(\C,\C)\raw\Z$).

In the presence of a proper $G$ action, one uses the equivariant K-homology group $K_0^G(M)=KK^G(C_0(M),\C)$, which is defined like $KK(C_0(M),\C)$, but with the
additional stipulation that the Hilbert spaces $H^{\pm}$ in the Kasparov bimodule
$(F,H^+,H^-)$ are unitary $G$ modules, 
in such a way that $F$ is equivariant,  
and  the \rep s of $C_0(M)$ on $E^{\pm}$ are covariant under $G$
\cite{Kas2,Val}. One now has a canonical map $K_0^G(M)\raw K_0(C^*(G))$, called
the analytic assembly map, which plays a key role in the  Baum--Connes
conjecture \cite{BCH}. Replacing $K_0(C^*(G))$ with $KK(\C,C^*(G))$, 
this map is defined by a  slight generalization of the
 construction of the element $[\hat{\DS}\mbox{}^+,E^+(\CS),E^-(\CS)]$ of 
$KK(\C,C^*(G))$ explained prior to (\ref{Q1}); cf.\ \cite{Val} for details. 
The basic idea is to define the $C_c(G)$-valued inner products (\ref{Gip})
on the dense subspace $C_c(M)L^2(\CS^{\pm})$,  completing these
subspaces in the norm (\ref{norm}) to obtain the  Hilbert $C^*(G)$ modules $E^{\pm}(\CS)$.\footnote{We here assume that $G$ is unimodular, which guarantees that (\ref{Gip}) is positive.
This was shown for discrete $G$ in Lemma 3 in \cite{Val03}, but the proof
apparently works  for unimodular groups in general. In general, the construction in the preceding section produces a Hilbert module over the reduced group \ca\ $C_r^*(G)$ \cite{BCH}. This is sufficient for the Baum--Connes conjecture, but not for our generalized Guillemin--Sternberg conjecture.}

 It follows that the element of $KK(\C,C^*(G))$ that quantizes
the dual pair $pt\law M\raw \g^*_-$ a la Bott is just the image of the pertinent fundamental class of $M$ under the analytic assembly map.\footnote{Cf.\ \cite{NPLHaag}
for an exposition of the link between the analytic assembly map and 
$C^*$-algebraic deformation quantization, following Connes's discussion of this map in E-theory \cite{Con}.}
 The functoriality condition (\ref{GSrev}) remains well defined,
but the computation (\ref{Q3}) is invalid for noncompact groups, so that 
for noncompact $G$ the left-hand side of the  Guillemin--Sternberg conjecture is simply given by the left-hand side instead of the right-hand side of (\ref{Q3}).\footnote{A complication
arises when $M$ does not admit a $G$ invariant $\spinc$ structure. For techniques to overcome this cf.\ \cite{HS,Par2}.}
This yields a generalization of the Guillemin--Sternberg conjecture to noncompact groups, where $\Gindex(\DS^+)_0$ 
in  (\ref{G0}) is now reinterpreted
as the image of $\Gindex(\DS^+)\in K_0(C^*(G))$ under the map
$K_0(C^*(G))\raw \Z$ induced in K-theory by the $\mbox{}^*$-homomorphism $f\mapsto \int_G dx\, f(x)$ from $C^*(G)$ to $\C$. 
 
As a first example, consider the case where $G=\Gamma$ is discrete and infinite. One then simply has $M^0=M/\Gm$, and $\DS^+_0$ is just the operator on $M/\Gm$ whose lift is $\DS^+$. Using Atiyah's $L^2$-index theorem \cite{AtBour}, our generalized \GSC\  is equivalent to
$$ \Gindex(\DS^+)_0=\mathrm{tr}\circ\pi_*\circ \Gindex(\DS^+). $$
Here $\pi_*:  K_0(C^*(\Gm))\raw  K_0(C_r^*(\Gm))$ is the K-theory map
functorially induced by the canonical projection $\pi: C^*(\Gm)\raw C^*_r(\Gm)$, and $\mathrm{tr}: K_0(C^*(\Gm))\raw \mathbb{C}$ is defined by the pairing of the trace $f\mapsto f(e)$ on $C^*_r(\Gm)$ (seen as a cyclic cocycle) with K-theory \cite{Con}.
  \section{Foliation theory and quantization}
A second  generalization of the Guillemin--Sternberg conjecture
arises when one considers strongly Hamiltonian actions of Lie groupoids on symplectic manifolds; the pertinent  symplectic reduction procedure
was first studied in \cite{MiWe}, and is actually a special case of the product
$\otc$ \cite{BRW,NPL}. Furthermore,  the appropriate construction of elements of $K_0(C^*(G))$ has been given in \cite{Con,Pat}.
 A very interesting special case  comes from foliation theory, as follows (cf.\ \cite{C82,Con,HS,Moe,Mrcun}). Let $(V_i,F_i)$, $i=1,2$, be foliations with associated holonomy groupoids $G(V_i,F_i)$ (assumed to be Hausdorf for simplicity). A smooth generalized
map $f$ between the leaf spaces $V_1/F_1$ and $V_2/F_2$ is defined as a 
smooth right principal
bibundle $M_f$ between the Lie groupoids $G(V_1,F_1)$ and $G(V_2,F_2)$.
Classically, such a bibundle defines a dual pair $T^*F_1\law T^*M_f\raw T^*F_2$
\cite{NPLLMP}. 
Here  $TF_i\subset TV_i$ is the tangent bundle to the foliation $(V_i,F_i)$, whose dual bundle $T^*F_i$ has a canonical Poisson structure.\footnote{The best way to see this is to 
interpret $TF_i$ as the Lie algebroid of $G(V_i,F_i)$, and to pass to the canonical
Poisson structure on the dual bundle $A^*(G)$ to the Lie algebroid  $A(G)$ of any Lie groupoid $G$.}  Quantum mechanically, $f$ defines an element \cite{C82,HS}
$$ f_!\in KK(C^*(G(V_1,F_1)), C^*(G(V_2,F_2))).$$
In our functorial approach to quantization, $f_!$ is interpreted as the quantization 
of the dual pair $T^*F_1\law T^*M_f\raw T^*F_2$. The functoriality of quantization
  among  dual pairs of the same type  then follows from the computations in \cite{HS,NPLLMP}. 
The construction and functoriality of shriek maps in \cite{AS1,C82} is a special case of this,
in which the $V_i$ are both trivially foliated. 
  
 \end{document}